\documentclass[prb,aps,superscriptaddress,showpacs,twocolumn,amssymb,amsmath,amstext,amsfont,noshowkeys]{revtex4}
\usepackage{graphicx}
\usepackage{theorem}
\usepackage{dcolumn}
\usepackage{bm}

\newcommand{\SSS}{\scriptscriptstyle}

\newcommand{\Dd}{{\rm d}}

\begin{document}

\title{Vortex Nucleation Induced Phonon Radiation from a Moving Electron Bubble in Superfluid $^4$He}
\author{Dafei Jin}
\affiliation{Department of Physics, Brown University, Providence, Rhode Island 02912, USA}
\author{Wei Guo}
\affiliation{Department of Physics, Yale University, New Haven, Connecticut 06520, USA}
\date{\today}

\begin{abstract}

We construct an efficient zero-temperature semi-local density functional to
dynamically simulate an electron bubble passing through superfluid $^4$He under
various pressures and electric fields up to nanosecond timescale. Our simulated
drift velocity can be quantitatively compared to experiments particularly when
pressure approaches zero. We find that the high-speed bubble experiences
remarkable expansion and deformation before vortex nucleation occurs.
Accompanied by vortex-ring shedding, drastic surface vibration is generated
leading to intense phonon radiation into the liquid. The amount of energy
dissipated by these phonons is found to be greater than the amount carried away
solely by the vortex rings. These results may enrich our understanding about
the vortex nucleation induced energy dissipation in this fascinating system.

\end{abstract}

 \maketitle

\section{Introduction}

An electron injected into liquid helium forms a bubble due to the Pauli
exclusion between an excess electron and helium
atoms.\cite{SommerPRL1964,SpringettPR1967} The dissipation mechanisms of a
moving electron bubble in superfluid $^4$He has attracted considerable research
interests for many
years.\cite{MeyerPR1958,RayfieldPR1964,StrayerPRL1971,PhillipsPRL1974,BowleyPRL1980,NancolasNature1985,SchwarzPRA1973,QVHII1991,EILH2007}
Above 1~K, the moving bubble experiences a drag force from collisions with
thermally excited phonons and rotons.\cite{MeyerPR1958} Below 1~K, this drag
force becomes very small, and even a weak electric field can accelerate the
bubble to a high speed until some new dissipation mechanisms set in. In the
high pressure regime, roton emission dominates the dissipation owing to the
relatively low Landau velocity $v_{\SSS\text{L}}$ and the high vortex-ring
nucleation critical velocity
$v_{\text{c}}$.\cite{PhillipsPRL1974,BowleyPRL1980} In the low pressure regime,
vortex-ring nucleation plays the key role. A vortex ring can attach to the
bubble to form a bubble-ring complex if the electric field is not overly
strong; otherwise, successive vortex rings can be shed away from the bubble
surface.\cite{RayfieldPR1964,StrayerPRL1971,NancolasNature1985}

Pioneering simulations using the Gross-Pitaevskii equation (GPE) have
demonstrated the above vortex-ring nucleation, trapping, and shedding
scenario.\cite{FrischPRL1992,BerloffPRB2000} But it is known to be difficult
for the efficient local GPE to reproduce helium
properties.\cite{BerloffJPA1999} In contrast, accurate nonlocal density
functional (NLDF) theories have been extensively applied to quasiparticles,
vortices, and ions-related problems in liquid
helium.\cite{DalfovoPRB1995,GiacomazziPRB2003} But they usually require
prohibitively high computational cost for dynamic simulations. In this paper,
we first introduce our well constructed semi-local density functional (SLDF),
associated with an optimized numerical scheme, that can reconcile both the
physical accuracy and the computational efficiency in its applicable
regime.\cite{StringariPRB1987,JinJLTP2010} We then present our dynamic
simulation for a moving electron bubble in pure superfluid $^4$He under low
pressures and strong electric fields, where successive vortex-ring shedding is
indeed observed. However, we shall point out that although vortex nucleation
does trigger the dissipation, the major part of energy loss may come from its
induced phonon radiation via surface vibration, rather than purely the
shed-away vortex rings.

\section{Theoretical Formulation}

We formulate our problem in the framework of a zero-temperature SLDF theory.
The system free energy density $\mathcal{G}$ consists of the helium part, the
electron part, and the helium-electron interaction part,
\begin{equation}
\mathcal{G} = \mathcal{G}_{\text{He}} + \mathcal{G}_{\text{e}} +
\mathcal{G}_{\text{He-e}}.
\end{equation}
The helium part $\mathcal{G}_{\text{He}}$ takes the form of
\begin{equation}
\begin{split}
\mathcal{G}_{\text{He}} =\ & \frac{\hbar^2}{2m_{\text{He}}} |\nabla\psi|^2 -\mu \varrho \\
& + \frac{1}{2} g_{2} \varrho^2 + \frac{1}{3} g_{3} \varrho^3 + \frac{1}{4} g_{4} \varrho^4\\
& + \frac{1}{2} h_{2} |\nabla \varrho|^2 + \frac{1}{3} h_{3} \varrho
|\nabla \varrho|^2 + \frac{1}{4} h_{4} \varrho^2 |\nabla \varrho|^2,\\
\end{split}\label{SLDF}
\end{equation}
where $\psi$ is the macroscopic helium wavefunction and $\varrho\equiv|\psi|^2$
is the local helium number density. The electron part $\mathcal{G}_{\text{e}}$
takes the form of
\begin{equation}
\mathcal{G}_{\text{e}} = \frac{\hbar^2}{2m_{\text{e}}} |\nabla\phi|^2 -
e\mathcal{E}z \eta,
\end{equation}
where $\phi$ is the single electron wavefunction and $\eta\equiv|\phi|^2$ is
the local electron number density. The $|\nabla\psi|^2$ and $|\nabla\phi|^2$
terms above are the helium and electron kinetic energy densities, with
$m_{\text{He}}$ and $m_{\text{e}}$ being the helium and electron masses,
respectively. $\mu$ is the helium chemical potential controlled by a given
pressure $p$, which fixes the helium number density $\rho$ in bulk.
$\mathcal{E}$ is the applied electric field, which drives the electron to move
along $z$ direction. $g_2$, $g_3$, $g_4$ and $h_2$, $h_3$, $h_4$ are all
fitting parameters to be explained below. The helium-electron interaction part
$\mathcal{G}_{\text{He-e}}$ takes the form of a contact collision
\begin{equation}
\mathcal{G}_{\text{He-e}} =  f_1 \varrho \eta,
\end{equation}
in which $f_1$ is a fitting parameter chosen so as to produce a $1$~\AA\
scattering length and hence a $1.0456$~eV potential barrier for an electron to
tunnel into homogeneous helium at zero temperature and zero
pressure.\cite{SommerPRL1964,SpringettPR1967}

The polynomial function of $\varrho$ in the second line of Eq.~(\ref{SLDF}) is
the helium-helium interaction energy density in the form of two-particle,
three-particle and four-particle contact collisions. In the homogeneous case,
it gives the helium internal energy density
\begin{equation}
\varepsilon[\rho] = \frac{1}{2} g_2 \rho^2 + \frac{1}{3} g_3 \rho^3 +
\frac{1}{4} g_4 \rho^4,
\end{equation}
from which one can derive the equation of state
\begin{equation}
p[\rho] = \frac{1}{2} g_2 \rho^2 + \frac{2}{3} g_3 \rho^3 + \frac{3}{4} g_4
\rho^4,
\end{equation}
the chemical potential
\begin{equation}
\mu[\rho] = g_2 \rho + g_3 \rho^2 + g_4 \rho^3,
\end{equation}
and the sound velocity
\begin{equation}
c[\rho] = \sqrt{\left( g_2 \rho + 2 g_3 \rho^2 + 3 g_4 \rho^3 \right) /
m_{\text{He}} }.
\end{equation}
By choosing $g_2$, $g_3$ and $g_4$ according to the well-known Orsay-Trento
density functional,\cite{DalfovoPRB1995} the experimentally measured above
quantities at zero temperature can be very well produced.

The mixture polynomial function of $\varrho$ and $|\nabla \varrho|^2$ in the
third line of Eq.~(\ref{SLDF}) contains our introduced semi-local interactions
also up to four-particle collisions. They are the lowest-order
gradient-expansion corrections to the local interactions and are essential for
incorporating the helium surface tension into our theory
\begin{equation}
\begin{split}
\sigma[\rho] = & \int_0^\rho\Dd \varrho\ b[\varrho]\sqrt{\frac{\hbar^2}{8m_{\text{He}}\varrho } }\\
& \frac{ 2(\varepsilon[\varrho]-\varepsilon[\rho])-(\mu[\rho] +
\varepsilon[\rho]/\rho) (\varrho - \rho)}
{\sqrt{(\varepsilon[\varrho]-\varepsilon[\rho])-\mu[\rho](\varrho - \rho)}},
\end{split}\label{SurfaceTension}
\end{equation}
where the dimensionless function
\begin{equation}
b[\rho] = \sqrt{1+\frac{8m_{\text{He}} \rho}{\hbar^2} \left( \frac{1}{2} h_2 +
\frac{1}{3} h_3 \rho + \frac{1}{4} h_4 \rho^2 \right)}.
\end{equation}
By choosing $h_2$, $h_3$ and $h_4$ appropriately, the experimentally measured
surface tension at zero temperature and zero pressure can be produced. In
addition, Eq.~(\ref{SurfaceTension}) is a generalized formula to allow not only
zero pressure, when the liquid is in equilibrium with vacuum, but also a
positive pressure, when the liquid is in contact with an impenetrable wall. In
both situations, the liquid density drops from its bulk value $\rho$ to 0
within a thin interfacial layer. $\sigma[\rho]$ measures the energy change per
unit surface area due to this density bending, in comparison with the internal
energy per unit area held by the same number of particles in
bulk.\cite{StringariPRB1987} Such a generalized definition on the surface
tension turns out to be useful for the electron bubble problem, where the
bubble boundary pushes away the liquid basically like an impenetrable wall, and
so gives rise to a nontrivial surface energy under any finite
pressures.\cite{SpringettPR1967}

\begin{table}[htb]
\caption{The chosen fitting parameters of our SLDF.}\label{Parameter}
\begin{ruledtabular}
\begin{tabular}{crl}
 Parameter & Value\qquad\qquad & Unit \\
\hline
$f_1$ & $5.55671\times10^5$ & K~\AA$^3$ \\
$g_2$ & $-7.18990\times10^2$ & K~\AA$^3$ \\
$g_3$ & $-3.61779\times10^4$ & K~\AA$^6$ \\
$g_4$ & $2.47799\times10^6$ & K~\AA$^9$ \\
$h_2$ & $1.16950\times10^4$ & K~\AA$^5$ \\
$h_3$ & $-1.35048\times10^6$ & K~\AA$^8$ \\
$h_4$ & $3.46549\times10^7$ & K~\AA$^{11}$
\end{tabular}
\end{ruledtabular}
\end{table}

\begin{figure}[htb]
\includegraphics[scale=0.8]{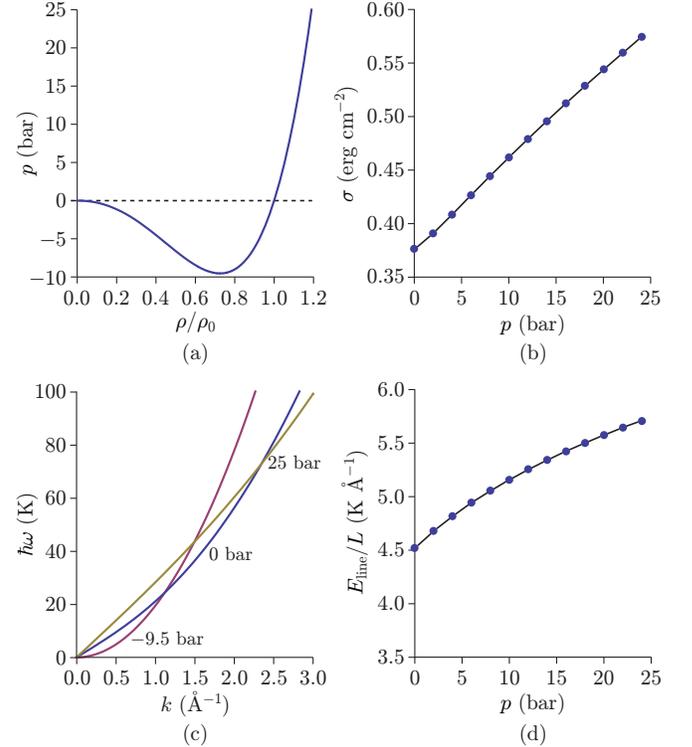}
\caption{(color online). The helium properties in our SLDF. (a) The equation of
state. (b) The surface tension versus pressure. (c) The excitation spectrum
under the spinodal pressure $-9.5$~bar, the saturation pressure $0$~bar, and
the melting pressure $25$~bar. (d) The vortex-line energy per unit length
versus pressure with a 20~\AA\ cutoff radial distance.}
\label{HeliumProperties}
\end{figure}

Table~\ref{Parameter} lists all the chosen fitting parameters of our SLDF.
Figure \ref{HeliumProperties}(a) shows the equation of state. At $p=0$~bar,
$\rho_0=0.021836$~\AA$^{-3}$, $\mu_0=-7.1500$~K, and $c_0=237.70$~m~s$^{-1}$.
Figure \ref{HeliumProperties}(b) shows the surface tension versus pressure. At
$p=0$~bar, $\sigma_0 = 0.37554$~erg~cm$^{-2}$ with a calculated 5.8~\AA\
surface thickness (defined by 10\%~to~90\% helium density) consistent with
experiments and other theories.
\cite{DalfovoPRB1995,StringariPRB1987,PenanenPRB2000}

The excitation spectrum in our theory can be found to be of the Bogoliubov type
\begin{equation}
\hbar^2\omega^2 =  c^2[\rho] \hbar^2 k^2 + b^2[\rho] \left(\frac{\hbar^2 k^2
}{2m_{\text{He}}}\right)^2,\label{Excitation}
\end{equation}
whose pressure dependence is enclosed in the sound velocity $c[\rho]$ from the
local interactions, and the dimensionless function $b[\rho]$ from the
semi-local interactions. Compared with the efficient local GPE, although our
SLDF can indeed produce more realistic helium bulk and surface properties, and
meanwhile maintain the same computational efficiency, it still cannot
incorporate the backflow effect and the roton excitation unless some nonlocal
interactions are introduced.\cite{BerloffJPA1999,DalfovoPRB1995}
Figure~\ref{HeliumProperties}(c) shows the excitation spectrum under three
typical pressures: the spinodal pressure $-9.5$~bar, the saturation pressure
$0$~bar, and the melting pressure $25$~bar. The primary reason for employing
three semi-local parameters, $h_2$, $h_3$ and $h_4$, in our SLDF instead of a
single one, as that done in the well-known Stringari-Treiner density
functional,\cite{StringariPRB1987} is to remove some pathological behavior of
the excitation spectrum. A single semi-local parameter being used to fit the
surface tension at zero pressure will result in too large of a $b[\rho]$ and
hence unrealistically high excitation energies even at just moderate
momentums.\cite{JinJLTP2010} This will make the liquid overrigid against
density fluctuation and may bring on artifacts in dynamic simulations. With the
three semi-local parameters, we can keep $b[\rho]$ sufficiently small, and also
let it slowly decrease with pressure to imitate the tendency of roton gap with
pressure in real helium.

We can also find in our theory the rectilinear vortex-line energy per unit
length, by calculating the radial number density distribution $\varrho(r)$
under a given pressure,
\begin{equation}
\begin{split}
\frac{E_{\text{line}}}{L} = &\int 2\pi r \Dd r \left( \frac{\hbar^2}{2m_{\text{He}}r^2}\varrho + p -\mu \varrho \right. \\
 & + \frac{1}{2} g_{2} \varrho^2 + \frac{1}{3} g_{3} \varrho^3 + \frac{1}{4} g_{4}
 \varrho^4 + \frac{\hbar^2}{8m_{\text{He}}\varrho} \varrho'^2 \\
 & \left. + \frac{1}{2} h_{2} \varrho'^2 + \frac{1}{3} h_{3} \varrho \varrho'^2 +
 \frac{1}{4} h_{4} \varrho^2 \varrho'^2 \right),
\end{split}
\label{VortexEnergy}
\end{equation}
where $\varrho$ drops from its bulk value $\rho$ to 0 as $r\rightarrow0$, and
$\varrho'\equiv\Dd\varrho/\Dd r$. The first term in the first line contributes
to the kinetic energy $K_{\text{line}}/L$ due to one unit of quantum
circulation. All the other terms contribute to the potential energy
$U_{\text{line}}/L$ due to the density bending.\cite{MarisJLTP1994} The
critical negative pressure for the free expansion of a vortex core is found to
be close to $-7$~bar. Figure~\ref{HeliumProperties}(d) shows the vortex-line
energy per unit length versus pressure with a $d=20$~\AA\ cutoff radial
distance. At $p=0$~bar, we find $K_{\text{line}}/L=2.94$~K~\AA$^{-1}$ and
$U_{\text{line}}/L=1.57$~K~\AA$^{-1}$. Putting them into the hollow-core model
\cite{QVHII1991}
\begin{equation}
\begin{split}
\frac{K_{\text{line}}}{L} &= \frac{\pi\hbar^2}{m_{\text{He}}} \rho_0 \ln\frac{d}{a}, \\
\frac{U_{\text{line}}}{L} &= \frac{\pi\hbar^2}{m_{\text{He}}} \rho_0 \delta,
\end{split}
\end{equation}
we can get the core size $a=0.581$~\AA\ and the core parameter $\delta=1.893$,
which imply a rather rigid core.

\section{Dynamic Simulation}

We perform the dynamic simulation in the experimental low pressure regime
$p=0\sim10$~bar and strong electric field regime
$\mathcal{E}=1\sim30$~MV~m$^{-1}$. Under these conditions, a high-speed
electron bubble can nucleate vortex rings. As is well-known, there are two
classic models on the formation of a vortex ring and bubble-ring
complex.\cite{SchwarzPRA1973,QVHII1991,EILH2007} One is the encircling-ring
(quantum transition) model, in which a full ring axisymmetrically appears on
the bubble equator first, then moves sideways and captures the bubble on its
core. The other one is the pinned-ring (peeling) model, in which a small
proto-ring non-axisymmetrically appears on the bubble equator first, then grows
up and retains the bubble on its core. Experimental evidence suggests that the
encircling-ring model gives a better description at lower temperatures in
isotopically purified $^4\mathrm{He}$, whereas the pinned-ring model gives a
better description at higher temperatures in natural $^4\mathrm{He}$, with the
existence of thermal rotons or $^3$He
impurities.\cite{RayfieldPR1964,StrayerPRL1971,BowleyPRL1980} For an already
formed bubble-ring complex, the strength of the electric field determines its
subsequent behavior. In a field of the order of $10^3\sim10^4$~V~m$^{-1}$, the
ring can keep trapping the bubble and may grow to micron size, which
significantly hinders the bubble motion.\cite{RayfieldPR1964} But in a field of
the order of $10^6\sim10^7$~V~m$^{-1}$, owing to the vanishing escape barrier,
the ring should separate from the bubble during several picoseconds with its
size almost unchanged. The bubble can then move more freely, as well as
successively shed away vortex rings.\cite{NancolasNature1985} Since our
simulation is for pure $^4\mathrm{He}$ at zero temperature, the axisymmetric
vortex-ring nucleation in the encircling-ring model is more suitable.
Furthermore, since our electric fields lie in the strong regime, the transient
non-axisymmetric capture-and-escape processes should have little influence on
the longtime physics. Thus we reduce the original three-dimensional problem
into two-dimensional by taking cylindrical symmetry with transverse coordinate
$r$ and longitudinal coordinate $z$. This simplification makes the numerical
integration up to nanosecond timescale computationally practical.

The coupled equations of motion for the macroscopic helium wavefunction $\psi$
and the single electron wavefunction $\phi$ can be derived from our SLDF as
\begin{equation}
\begin{split}
\dot{\psi} =& -\frac{i}{\hbar}\left[-\frac{\hbar^2}{2m_{\text{He}}}\nabla^2 - \mu + f_1|\phi|^2 \right.\\
& + g_2|\psi|^2 + g_3|\psi|^4 + g_4|\psi|^6 \\
& - \left(\frac{1}{3}h_3+\frac{1}{2}h_4|\psi|^2\right)\left(\nabla |\psi|^2 \cdot \nabla|\psi|^2\right) \\
& \left. - \left(h_2+\frac{2}{3}h_3|\psi|^2+\frac{1}{2}h_4|\psi|^4\right)
\nabla^2 |\psi|^2
\right]\psi,\\
\dot{\phi} =& -\frac{i}{\hbar}\left[-\frac{\hbar^2}{2m_{\text{e}}}\nabla^2 -
e\mathcal{E}z + f_1|\psi|^2 \right]\phi.
\end{split}
\label{EOM}
\end{equation}
The fourth-order finite-difference method in space and the fourth-order
Runge-Kutta method in time are used to do the numerical integration. The
computation grid on the $r$-$z$ plane is $500\times1000$ with the space step of
0.2~\AA. So the physical space is a pipe of 200~\AA\ in diameter and 200~\AA\
in length. We let it move along to keep the electron in the center. Systematic
numerical tests have been done to ensure that 0.2~\AA\ spatial resolution
provides enough precision for our problem, even when the vortex core structure
is involved.

Due to the big difference between the helium and electron intrinsic timescales
of the order of $m_{\text{He}}/m_{\text{e}}\sim7300$, we adopt the adiabatic
approximation. For every instantaneous helium configuration, we first find the
electron ground state by evolving the electron with 50 imaginary time steps of
0.001~fs, then develop helium with 1 real time step of 0.001~ps. To avoid the
outgoing sound waves from the central region being reflected on the grid
boundary, we set a space-dependent damping coefficient by replacing $i$ with
$\eta(s)+i$ in the helium equation of motion, where
\begin{equation}
\eta(s) = \frac{\bar{\eta}}{2} \left[
1+\tanh\left(\frac{s-\bar{s}}{\bar{w}}\right) \right],\
(s\equiv\sqrt{r^2+z^2}).
\end{equation}
We simply choose $\bar{s}=80$~\AA, $\bar{w}=1.0$~\AA, and $\bar{\eta}=0.5$. The
sound waves can travel freely through the undamped region ($s<\bar{s}$) but are
quickly attenuated in the damped region ($s>\bar{s}$). Physically, this looks
as if there is a strong zero-temperature heat sink surrounding the system.

\begin{figure}[hbt]
\includegraphics[scale=0.8]{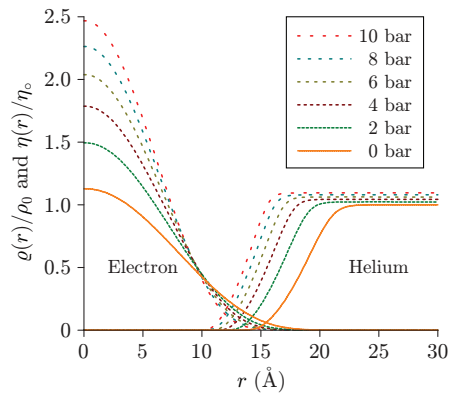}
\caption{(color online). The radial number density profiles of helium and
electron for a spherical static electron bubble under various pressures.}
\label{DensityProfiles}
\end{figure}

Figure~\ref{DensityProfiles} shows the calculated radial number density
profiles of helium $\varrho(r)/\rho_0$ and electron $\eta(r)/\eta_{\circ}$ for
a spherical static electron bubble under various pressures. Here
$\eta_{\circ}\equiv\pi/2R_{\circ}^3$ with $R_{\circ}\equiv20$~\AA\ being merely
a reference radius. The static bubble radius $R_{\text{b}}$ (defined at 50\%
helium density) can be found to be 18.52~\AA\ at $p=0$~bar and squeezed down to
13.50~\AA\ at $p=10$~bar, consistent with experiments and other
theories.\cite{SommerPRL1964,SpringettPR1967} We set these ground states as the
initial states to start the time evolution. For every single run with a
specified pressure and electric field, 8-core parallel computation at 2.83~GHz
was performed. Multiple runs with different pressures and electric fields were
conducted simultaneously on our computing cluster which possesses hundreds of
cores. Typically, it took about one month to reach 10~ns physical time.

\begin{figure*}[htb]
\centerline{\includegraphics[scale=0.8]{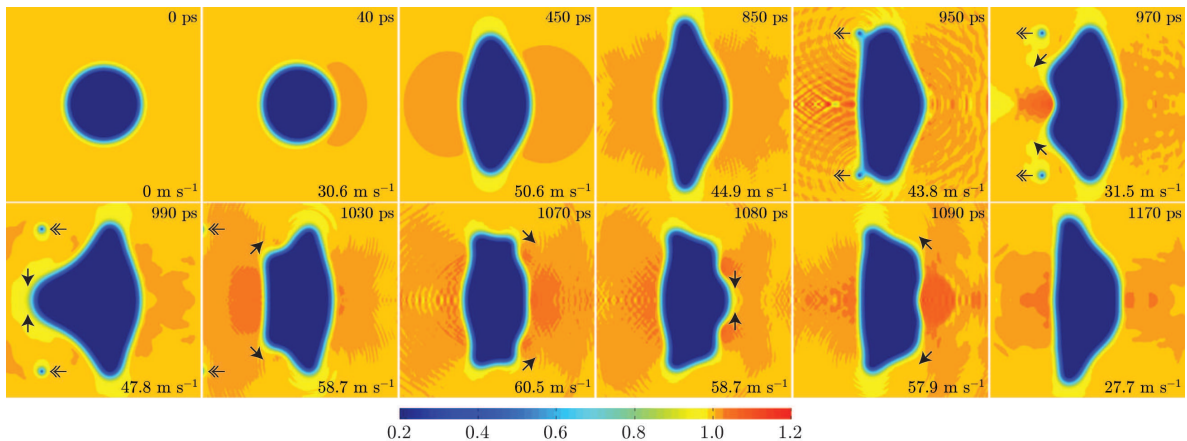}} \caption{(color online). The
snapshots of the helium number density profile around the moving electron
bubble at a sequence of times under zero pressure and 10~MV~m$^{-1}$ electric
field.\cite{Video} The bubble travels from left to right along $z$-direction
(the longitudinal direction). Each image corresponds to a physical size of
$100~\text{\AA}~\times~100~\text{\AA}$ and is reflected up and down in
$r$-direction (the transverse direction) justified by the imposed cylindrical
symmetry. The instantaneous velocity is given for each time. The open and solid
arrows indicate the movements of vortex ring and solitary ripplon relative to
the bubble as mentioned in the text.}\label{DissipationProcessD}
\end{figure*}

Figure~\ref{DissipationProcessD} shows the snapshots of the helium number
density profile $\varrho(r,z;t)/\rho_0$ around the moving electron bubble at a
sequence of times under $p=0$~bar and $\mathcal{E}=10$~MV~m$^{-1}$. The
instantaneous velocity $v$ at each time is approximated by linearly fitting the
displacement of the electron wavefunction-maximum over a $20$~ps time interval.
At $t=0$~ps, the bubble is on its spherical ground state of radius
$R_{\text{b}}=18.52$~\AA, when the driving field is just switched on. The
initial acceleration of the bubble has the value that is expected based on the
applied force $e\mathcal{E}$ and the hydrodynamic mass $2\pi R_{\text{b}}^3
m_{\text{He}}\rho_0 /3$. Up until $t=40$~ps, $v$ is already as high as
30.6~m~s$^{-1}$, whereas the bubble shape is only slightly squeezed and the
liquid in front is slightly compressed. As the bubble velocity increases
further, its volume keeps expanding and its shape keeps deforming into an
oblate spheroid due to the negative pressure on the bubble waist caused by the
Bernoulli effect.\cite{GuoAIPCP2006} The acceleration decreases remarkably
because of the increase in the effective mass so that $v$ approaches a constant
value about 50.6~m~s$^{-1}$ and the flow pattern exhibits quite well
fore-and-aft symmetry as shown at $t=450$~ps. After this stage, a narrow edge
girdling the bubble waist gradually sticks out signifying the onset of
vortex-ring nucleation. This adds a considerable inertia to the bubble motion,
making $v$ decrease to 44.9~m~s$^{-1}$ at $t=850$~ps, when the bubble attains
its largest equatorial radius about 44~\AA\ and its sharpest edge radius of
curvature about $4$~\AA. We may call the bubble velocity for the first
appearance of this characteristic behavior as the vortex-ring nucleation
critical velocity $v_{\text{c}}=44.9$~m~s$^{-1}$ under the prescribed
conditions. There is no dissipation before, since the electric work mainly
transforms to the fluid kinetic energy. But around this time the liquid density
starts to fluctuate accompanied by some phonon generation, presumably because
the local flow near the bubble waist becomes supersonic and keeps hitting the
nucleated vortex core.\cite{FrischPRL1992,BerloffPRB2000} At $t=950$~ps, a
vortex ring of radius $R_{\text{ring}}\approx 36$~\AA\ is detaching from the
bubble surface, while $v$ is 43.8~m~s$^{-1}$ and continues to slow down due to
the attraction from the vortex ring. The phonon generation appears more
intense, particularly peaked at a wavelength of the order of 10~\AA. The
unexpected large ring radius, almost twice as big as the static bubble radius,
is a unique observation of our dynamic SLDF simulation.

In the later stage, the vortex ring falls behind, leaving a drastic surface
vibration and velocity oscillation to the bubble in the next hundreds of
picoseconds. At $t=970$~ps, an annular surface protrusion like a solitary
ripplon is formed and sliding backwards, while $v$ is dragged down to
31.5~m~s$^{-1}$. This solitary ripplon then shrinks on the bubble tail at
$t=990$~ps, and radiates a large number of phonons backwards into the liquid as
shown at $t=1030$~ps. This provides a strong forward impulse to speed up the
bubble so that $v$ can even overshoot to 60.5~m~s$^{-1}$ at $t=1070$~ps. During
this time, the solitary ripplon surviving from the last collision bounces back
and propagates forwards. Around $ t=1080 \sim 1090$~ps, it shrinks on the
bubble head and deposits a large number of phonons forwards into the liquid.
This provides a strong backward impulse decelerating the bubble velocity to as
low as 27.7~m~s$^{-1}$ at $t=1170$~ps. Afterwards, some remnant surface
vibration and phonon radiation keep happening back and forth, while the bubble
velocity oscillatorily increases until a second vortex ring completes its
nucleation. Then all the processes repeat.

\section{Discussion}

\begin{figure}[hbt]
\includegraphics[scale=0.8]{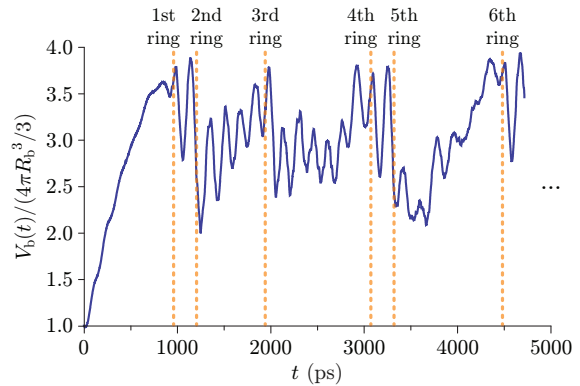}
\caption{(color online). The time evolution of the moving electron bubble
volume under zero pressure and 10~MV~m$^{-1}$ electric field. The dotted lines
mark the moments of the vortex-ring shedding events.} \label{BubbleVolume}
\end{figure}

Figure~\ref{BubbleVolume} shows the time evolution of the moving electron
bubble volume $V_{\text{b}}(t)/(4\pi R_{\text{b}}^3/3)$ under $p=0$~bar and
$\mathcal{E}=10$~MV~m$^{-1}$, obtained from the same evolution shown in
Fig.~\ref{DissipationProcessD} but extended to about 5~ns physical time. Here
$4\pi R_{\text{b}}^3/3$ with $R_{\text{b}}=18.52$~\AA\ is the spherical static
bubble volume at $t=0$~ps. As can be seen, before the dissipation processes
arise at $t\approx850$~ps, $V_{\text{b}}(t)$ keeps growing to about 3.6 times
the static volume. After that, $V_{\text{b}}(t)$ oscillates violently between
2.0 and 4.0 (average about 3.0) times the static volume. It approximately
implicates the hydrodynamic mass variation, and coincides with the bubble
velocity variation described above. The dotted lines mark the moments of the
vortex-ring shedding events. During $t=800\sim5000$~ps, there are six such
events in total. Although they do not exhibit a regular periodicity, we may
still roughly estimate the average time interval between two successive events
$\tau\approx700$~ps. Each time when a ring leaves the bubble, there follows a
huge volume change immediately and perhaps a series of small volume changes
later on. This can be related to the complex surface motion displayed in
Fig.~\ref{DissipationProcessD}.

\begin{figure}[htb]
\includegraphics[scale=0.8]{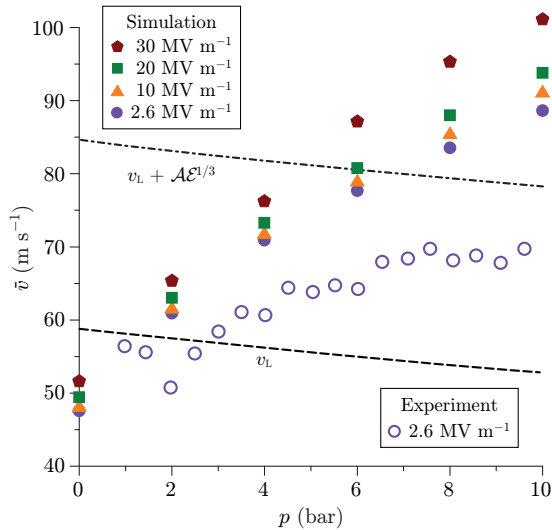}
\caption{(color online). The drift velocity versus pressure under 2.6, 10, 20,
30~MV~m$^{-1}$ electric fields from our simulation in comparison with the low
pressure part of experiment by Nancolas \textit{et. al.} at 0.3~K under
2.6~MV~m$^{-1}$. The dashed curve gives the Landau velocity for roton emission.
The dashed-dotted curve gives the theoretically extrapolated drift velocity
from 25~bar under 2.6~MV~m$^{-1}$ if roton emission was the only dissipation
mechanism.\cite{NancolasNature1985}} \label{VelocityPressure}
\end{figure}

Figure~\ref{VelocityPressure} shows the drift velocity $\bar{v}$ versus
pressure under $\mathcal{E}=2.6$, 10, 20, 30~MV~m$^{-1}$ from our simulation in
comparison with the low pressure part of experiment by Nancolas \textit{et.
al.} at 0.3~K under 2.6~MV~m$^{-1}$.\cite{NancolasNature1985} Our $\bar{v}$ is
calculated by linearly fitting the displacement of the electron
wavefunction-maximum over 2~ns after the bubble motion has become strongly
dissipative. Although the instantaneous velocity $v$ mentioned above undergoes
large oscillation, $\bar{v}$ is nearly constant depending only on the pressure
and electric field for a long time. According to Nancolas \textit{et. al.}, if
roton emission was the only dissipation mechanism, then in the limit
$\mathcal{E}\rightarrow0$~V~m$^{-1}$, $\bar{v}$ should follow the dashed curve
of Landau velocity $v_{\SSS\text{L}}$; whereas for
$\mathcal{E}=2.6$~MV~m$^{-1}$, $\bar{v}$ should follow the dashed-dotted curve
based on the well-known relation $\bar{v}=v_{\SSS\text{L}} + \mathcal{A}
\mathcal{E}^{1/3}$, in which $\mathcal{A}$ takes its value at 25~bar with the
pressure dependence ignored. However, the experimentally measured $\bar{v}$
shows a significant drop from this curve, indicating an additional dissipation
mechanism in this low pressure regime believed to be vortex-ring nucleation.
Our work supports this general scenario especially for $p<3$~bar where the
simulation and experimental results quantitatively agree with each other. This
suggests that our SLDF can be an efficient and reliable tool to study
interesting dynamic problems, particularly under the natural experimental
condition $p\rightarrow0$~bar close to zero temperature. The lack of roton
excitation in our SLDF may not be problematic for two reasons that essentially
prohibit roton emission in this range. First, the Landau velocity
$v_{\SSS\text{L}}\approx 58$~m~s$^{-1}$ is substantially higher than the
vortex-ring nucleation critical velocity $v_{\text{c}}\approx 45$~m~s$^{-1}$.
Second, there is experimental evidence that $\mathcal{A}$ is divergent as
$p\rightarrow3$~bar from above.\cite{NancolasNature1985} For $p>3$~bar, our
simulation still exhibits a qualitatively correct behavior in terms of the
increasing $\bar{v}$ with increasing $p$ commonly understood as the inverse
proportionality of $v_{\text{c}}$ with respect to the bubble
size.\cite{SchwarzPRA1973,QVHII1991,EILH2007,FrischPRL1992} The relatively
large deviation between simulation and experiment is presumably caused by the
lack of roton excitation in our SLDF, whereas in reality vortex-ring nucleation
and roton emission coexist in this pressure range.

Although in Fig.~\ref{DissipationProcessD} we have presented that vortex
nucleation is indeed the origin of dissipation, it is nontrivial to investigate
how the energy is actually taken away from the system. While this is normally
attributed to purely the frequent vortex-ring shedding, our analysis suggests
otherwise. The energy associated with a large single vortex ring at zero
pressure can be estimated by \cite{QVHII1991}
\begin{equation}
E_{\text{ring}} = \frac{\pi\hbar^2}{m_{\text{He}}} \rho_0 2\pi R_{\text{ring}}
\left(\ln\frac{8R_{\text{ring}}}{a}-2+\delta\right).
\end{equation}
Based on our preceding calculation about the vortex properties in our density
functional, when $R_{\text{ring}}\approx36$~\AA\ as shown in
Fig.~\ref{DissipationProcessD}, we have $E_{\text{ring}}\approx1147$~K. As a
result, the average rate of energy loss to the vortex rings is
$E_{\text{ring}}/\tau \approx 1147~\text{K}~/~700~\text{ps} = 2.3
\times10^{-11}~\text{W}$. On the other hand, the average rate of energy input
to the system from the electric work is $ e\mathcal{E} \bar{v} \approx
10~\text{MeV~m}^{-1}\times48~\text{m~s}^{-1}= 7.7\times10^{-11}~\text{W}$.
Strikingly, the released vortex rings only take away about 30\% of the total
input energy. According to the processes illustrated in
Fig.~\ref{DissipationProcessD} and discussed above, the remaining energy loss
can be primarily attributed to the phonon radiation through surface vibration
persistently occurring in the time interval between successive vortex-ring
shedding events. Under a fixed electric field with an increasing pressure, we
even notice a tendency for such a dissipation channel to become increasingly
dominant.

\section{Conclusion}

We have carried out dynamic simulation for a moving electron bubble in pure
superfluid $^4$He under low pressures and strong electric fields, using a
zero-temperature semi-local density functional theory that embodies realistic
$^4$He bulk and surface properties. It visualizes the underlying microscopic
processes from picosecond to nanosecond timescale, which cannot yet be directly
observed through any experimental apparatus today. Not only have we confirmed
that theory and experiment show good agreement on the bubble mobility
approaching zero pressure, we have also discovered that the pronounced surface
vibration and phonon radiation as a result of vortex nucleation dissipate more
energy than the shed-away vortex rings alone. These results may enrich our
understanding about the vortex nucleation induced energy dissipation in this
fascinating system and so warrant further experimental and theoretical studies.

\begin{acknowledgements}

The authors are very grateful to Professors H.J.~Maris, F. Ancilotto,
M.~Barranco, M.~Pi, and D.~Mateo for helpful discussions. This work was
supported in part by the National Science Foundation through Grant No.
DMR-0605355.

\end{acknowledgements}

\end{document}